\documentclass[12pt]{article}
\usepackage{cite}
\def\be#1{\begin{equation}\label{#1}}

          \def\ee{\end{equation}}

\def\pa{\partial}
\def\equ#1{(\ref{#1})}

\title{On the zero modes of the Faddev-Popov operator in the Landau gauge}
\author{R. R. Landim$^{a,b}$\footnote{email: renan@fisica.ufc.br} , L. C. Q.
Vilar$^{a}$\footnote{email: lcqvilar@gmail.com} , O. S. Ventura$^{c}$\footnote{email: ozemar.ventura@cefet-rj.br} , V.
E. R. Lemes$^{a}$\footnote{email: vitor@dft.if.uerj.br}\\
 \small \em $^a$Instituto de F\'\i sica, Universidade do Estado do Rio de Janeiro,\\
\small \em Rua S\~{a}o Francisco Xavier 524, Maracan\~{a}, Rio de Janeiro - RJ, 20550-013, Brazil\\
 \small \em $^b$Departamento de F\'{\i}sica, Universidade Federal do Cear\'{a}\\
\small \em Caixa Postal 6030, Campus do Pici, 60455-760, Fortaleza - Cear\'{a},
Brazil\\
\small \em $^c$Centro Federal de Educa\c{c}\~ao Tecnol\'ogica do Rio de
Janeiro\\
\small\em Av.Maracan\~a 249, 20271-110, Rio de Janeiro - RJ, Brazil}
\begin{document}
\maketitle
\begin{abstract}
Following Henyey procedure \cite{Henyey:1978qd},  we construct examples of zero modes of the Faddev-Popov operator in the Landau gauge in Euclidean space in $D$ dimensions, for both $SU(2)$ and $SU(3)$ groups. We consider gauge field configurations $A^a_\mu$ which give rise to a field strength, $F^a_{\mu\nu} =\partial_\mu A^a_\nu -\partial_\nu A^a_\mu + f^{abc}A^b_\mu A^c_\nu$,  whose nonlinear term,  $ f^{abc}A^b_\mu A^c_\nu$, turns out to be nonvanishing.   To our knowledge, this is the first time where such a non-abelian  configuration is explicitly obtained in the case of $SU(3)$
in $4D$.
\end{abstract}

\section{Introduction}
The Gribov problem is by now an old issue in theoretical physics. Its origin
can be traced back to the seminal paper of V. N. Gribov in the late
‘70s\cite{Gribov:1977wm}.
And the sense of this problem is of a rather simple nature, at least in its
mathematical meaning: the usual gauge fixations used in quantum field
theories do not in general completely extinguish the gauge freedom. Let
us take, as an example, the Landau gauge

\be{landau0}
\partial_\mu A^a_\mu=0.
\ee

If we start from a gauge connection $A^a_\mu$ satisfying \equ{landau0}, we can
ask if a gauge transformation with gauge parameter $\omega^a$,
\be{gauge1}
A'^a_\mu = A^a_\mu+D_\mu\omega^a ,
\ee
could lead to a new connection $A'^a_\mu$ satisfying the Landau condition
\equ{landau0}.
This, in fact, imposes a constraint to the gauge parameter

\be{zeromode0}
\partial^2\omega^a+f^{abc}A_\mu^b\partial_\mu\omega^c=0.
\ee

Then, once we admit that this set of equations can be solved for $\omega^a$
we assure that the field theory is still plagued with gauge copies. And if
we believe that this is troublesome, we need to pursue a more stringent
procedure than Landau gauge fixing in order to forbid such freedom.
As this problem is a general feature of any standard gauge fixing
\cite{Singer:1978dk}, we understand
that a whole new approach to quantum field theory is required. This
program was established by D. Zwanziger
\cite{Zwanziger:1988jt,Zwanziger:1989mf},  and it has found
recent developments along a BRST point of view
\cite{Capri:2010hb,Gomez:2009tj,Dudal:2010hj,
Sorella:2009vt}.
Among those, we call attention to the refined Gribov propagator  for
the gauge particles of the Gribov-Zwanziger theory \cite{Dudal:2008sp}

\begin{eqnarray}
   \langle A^a_{\mu}(p) A^b_{\nu}(-p)\rangle =\frac{p^2 + M^2}{p^4
+ (M^2+m^2)p^2 + 2 g^2 N \gamma^4 + M^2 m^2
}\left[\delta_{\mu\nu} - \frac{p_{\mu}p_{\nu}}{p^2}
\right]\delta^{ab}, \label{gluonprop}
\end{eqnarray}
to the construction of new gauge observables\cite{Dudal:2010cd},
and to the reading of this effect as a symmetry breaking process
\cite{Vilar:2011sp}.

It is remarkable that new-found results from the Lattice
points\cite{Cucchieri:2011ig}  to gluon
propagators with the structure predicted in \equ{gluonprop}, showing that the
Gribov-Zwanziger theory can be an alternative route towards confinement.

Although the Gribov problem is well established from what we exposed up
to now, the question of the actual solutions of eqs. \equ{zeromode0}, which is the
source of the problem, has not been tackled in all its extent. For instance,
only particular SU(2) solutions can be found in the literature
\cite{Henyey:1978qd,Guimaraes:2011sf,Capri:2012ev} . Solutions for the most
important case of SU(3) are still missing. Even their existence has not been
settled yet. This is the gap we intend to fill in the end of this work, which is organized as follows: in Section $2$ we establish the mathematical problem starting from the zero modes eqs. \equ{zeromode0}; in Sections $3$ and $4$ we work out the cases of the Lie groups $SU(2)$ and $SU(3)$ respectively; and in our conclusion, we show the form of a non-abelian $4D$ $SU(3)$ gauge copy of the Gribov kind. An appendix is included to show the properties of the angular eigenfunctions used along the text.

\section{The zero modes}
Let us start by fixing the notations and conventions: color indices are positioned on top and space-time indices are at the bottom, and summation  of repeated indices is assumed. The dimension of the Euclidean space is $D\ge2$ and the gauge group is $SU(N)$ if not otherwise specified. We want to obtain the solution of the eq. \equ{zeromode0} in the Landau gauge \equ{landau0}. In this gauge we can write
\be{amu}
A_\mu^a(x)=M^a_{\mu\nu}(r)x_\nu,
\ee
where $r^2=x_\mu x_\mu$ is the radial coordinate and $M^a_{\mu\nu}(r)=-M^a_{\nu\mu}(r)$, in order to satisfy \equ{landau0}. By making a substitution of the particular solution \equ{amu} into eq. \equ{zeromode0}, we obtain
\be{zm1}
\partial^2\omega^a+f^{abc}M^b_{\mu\nu}(r)x_\nu\partial_\mu\omega^c=0.
\ee
The operator $M^a_{\mu\nu}(r)x_\nu\pa_\mu$ that acts in $\omega^c$ can be written  in terms of the  angular momentum operator $\hat{L}_{\mu\nu}=x_\mu\pa_\nu-x_\nu\pa_\mu$:
\be{hato}
\hat{O}^a=M^a_{\mu\nu}(r)x_\nu\pa_\mu=\frac{1}{2}M^a_{\mu\nu}(r)(x_\nu\pa_\mu-x_\mu\pa_\nu)=\frac{1}{2}M^a_{\mu\nu}(r)\hat{L}_{\nu\mu}.
\ee
We can show that in  $D$ dimensions (see the appendix A),
\be{lap}
\pa^2=\frac{1}{r^{D-1}}\frac{\pa}{\pa r}(r^{D-1}\frac{\pa}{\pa r})+\frac{1}{r^2}\hat{L}^2,
\ee
where $\hat{L}^2=\hat{L}_{\mu\nu}\hat{L}_{\mu\nu}/2$.

Since the operator $\hat{L}^2$ commutes with $\hat{L}_{\mu\nu}$ and $\hat{L}_{\mu\nu}f(r)=0$, this implies that $\hat{L}^2$ commutes with $\hat{O}^a$. We can try a solution of eq. \equ{zm1} in the form
\be{omega}
\omega^a\rightarrow\omega^a_l=\tau^a_{\mu_1\mu_2\cdots
\mu_l}(r)Q_{\mu_1\mu_2\cdots \mu_l}(\Omega),
\ee
where $Q_{\mu_1\mu_2\cdots \mu_l}(\Omega)$ are eigenfunctions of $\hat{L}^2$ for
a given $l$(see the appendix A):
\be{autof}
\hat{L}^2 Q_{\mu_1\mu_2\cdots \mu_l}(\Omega)=-l(l+D-2)Q_{\mu_1\mu_2\cdots
\mu_l}(\Omega).
\ee
Substuting into  eq. \equ{zm1} we obtain
\be{omega2}
\hat{S}_{lD}\tau^a_{\mu_1\mu_2\cdots \mu_l}(r)Q_{\mu_1\mu_2\cdots
\mu_l}(\Omega)+f^{abc}\tau^c_{\mu_1\mu_2\cdots
\mu_l}(r)\hat{O}^bQ_{\mu_1\mu_2\cdots \mu_l}(\Omega)=0.
\ee
where
\be{S}
\hat{S}_{lD}=\frac{1}{r^{D-1}}\frac{d}{dr}(r^{D-1}\frac{d}{dr})-\frac{l(l+D-2)}{
r^2}.
\ee
The action of the operator $\hat{O}^c$ on $Q$ is given by
\be{OA}
\hat{O}^cQ_{\mu_1\mu_2\cdots \mu_l}=M^c_{\mu_1\tau}(r)Q_{\tau\mu_2\cdots \mu_l}+
M^c_{\mu_2\tau}(r)Q_{\mu_1\tau\cdots
\mu_l}+\cdots+M^c_{\mu_l\tau}(r)Q_{\mu_1\mu_2\cdots \tau}.
\ee
For $l=0$, $\omega^a_0=\tau^a(r)$. The solution \equ{omega} is just
$\omega^a_0=\alpha^a+\beta^a/r^{D-2}$, with $\alpha^a$ and $\beta^a$ constants, and
$M^a_{\mu\nu}(r)$ in \equ{amu} becomes arbitrary. However, with this solution
the gauge transformation \equ{gauge1} leads to a field that
diverges for $r=0$ unless $\beta^a=0$. For $\beta^a=0$, we have a constant $\omega^a$.
This is a trivial solution. The next nontrivial possible solution can be derived
for $l=1$. In this case we have
$\omega^a(x)=\tau^a_\mu(r)Q_\mu(\Omega)$, where $Q_\mu(\Omega)=x_\mu/r$ and
$\hat{O}^c(r)Q_\mu=M^c_{\mu\nu}(r)Q_\nu$. Since the $x_\mu$ are linearly
independent, we have the equation for
$\tau^a_\mu(r)$:
\be{es1}
\hat{S}_{1D}\tau^a_{\mu}(r)-f^{abc}M^b_{\mu\nu}(r)\tau^c_\nu(r)=0.
\ee
Now, we can mention two different paths to obtain solutions of \equ{es1}: i) we
can assign functions for $\tau^a_\mu(r)$ and try to obtain $M^b_{\mu\nu}(r)$ or
ii) assign functions to $M^b_{\mu\nu}(r)$ and
solve the differential equation for $\tau^a_\mu(r)$. In the case ii), the set of
eqs. \equ{es1}  is a second order coupled differential system and to obtain
$\tau^a_\mu(r)$ becomes a rather tough task. In the
other case, we have an algebraic system with $(N^2-1)D$ equations with
$(N^2-1)D(D-1)/2$ unknowns $M^a_{\mu\nu}(r)$. We can have a unique solution when
the number of equations equals the number of
unknowns. This occurs in $D=3$ for any group. In $D=2$, the number of unknowns
is smaller than the number of equations for any group. In order to obtain
solutions in $D=2$, we must cancel by hand, consistently, some of the
$\tau^a_\mu$. When $D>3$, we find $(N^2-1)D(D-1)/2-(N^2-1)D=(N^2-1)D(D-3)/2$
degrees of freedom in the system. We can choose them to be zero, for example,
and obtain the
$(N^2-1)D$
remaining $M^a_{\mu\nu}$'s in terms of the $(N^2-1)D$ $\tau^a_\mu$'s. Actually,
this property will allow our strategy to obtain higher dimensional solutions
from the lower dimensional and simpler cases.

\section{$SU(2)$ group}
Now, we start looking for the solutions of \equ{es1} for the $SU(2)$ group. The
Lie-algebra valued gauge field is $A_\mu=A^a_\mu\sigma^a/2$, where the
$\sigma^a,~a=1,2,3,$ are the Pauli matrices obeying
$[\sigma^a,\sigma^b]=2i\epsilon^{abc}\sigma^c$. In $D=2$ the matrix
$M^a_{\mu\nu}(r)$ has three non-null
elements: $M^a_{12}=-M^a_{21}=g^a(r)$.  In this case, we have the equations:

\begin{eqnarray}\label{d2su2}
 \frac{1}{r}\frac{d}{dr}(r\frac{d\tau^1_1(r)}{dr})-\frac{\tau^1_1(r)}{r^2}
+g^3(r)\tau^2_2(r)-g^2(r)\tau^3_2(r)=0,\nonumber\\
 \frac{1}{r}\frac{d}{dr}(r\frac{d\tau^1_2(r)}{dr})-\frac{\tau^1_2(r)}{r^2}
-g^3(r)\tau^2_1(r)+g^2(r)\tau^3_1(r)=0,\nonumber\\
 \frac{1}{r}\frac{d}{dr}(r\frac{d\tau^2_1(r)}{dr})-\frac{\tau^2_1(r)}{r^2}
+g^1(r)\tau^3_2(r)-g^3(r)\tau^1_2(r)=0,\\
 \frac{1}{r}\frac{d}{dr}(r\frac{d\tau^2_2(r)}{dr})-\frac{\tau^2_2(r)}{r^2}
-g^1(r)\tau^3_1(r)+g^3(r)\tau^1_1(r)=0,\nonumber\\
 \frac{1}{r}\frac{d}{dr}(r\frac{d\tau^3_1(r)}{dr})-\frac{\tau^3_1(r)}{r^2}
-g^1(r)\tau^2_2(r)+g^2(r)\tau^1_2(r)=0,\nonumber\\
 \frac{1}{r}\frac{d}{dr}(r\frac{d\tau^3_2(r)}{dr})-\frac{\tau^3_2(r)}{r^2}
+g^1(r)\tau^2_1(r)-g^2(r)\tau^1_1(r)=0\nonumber,
\end{eqnarray}
A way to obtain a solution of \equ{d2su2}  is, for example,  choosing
$\tau^1_1(r)=\tau^2_2(r)=\tau(r)$, the remaining $\tau$'s equal to zero and
$g^1=g^2=0$ . Then, we obtain
\be{sol12d}
g^3(r)=-\frac{1}{\tau(r)}\left(\frac{1}{r}\frac{d}{dr}(r\frac{d\tau(r)}{dr}
)-\frac{\tau(r)}{r^2}\right).
\ee
Consequently, we have $\omega=(\tau(r)x_1/r,\tau(r)x_2/r,0)$,
$A_\mu^1=A_\mu^2=0$ and $A^3_1(x)=x_2 g^3(r)$, $A^3_2(x)=-x_1 g^3(r)$.

We can obtain solutions of this type for $D$ dimensions by restricting
$M^a_{\mu\nu}$
to $M^3_{12}=-M^3_{21}=g^3(r)$ and $M^a_{\mu\nu}=0$ for $i,j>2$ and $a\ne3$. In
this case we have solutions like \equ{sol12d}:
\be{sol1dd}
g^3(r)=-\frac{1}{\tau(r)}\left(\frac{1}{r^{D-1}}\frac{d}{dr}(r^{D-1}\frac{
d\tau(r)}{dr})-(D-1)\frac{\tau(r)}{r^2}\right),
\ee
and $\omega=(\tau(r)x_1/r,\tau(r)x_2/r,0,0,\cdots,0)$, $A_\mu^1=A_\mu^2=0$,
$A^3_1=x_2 g^3(r)$, $A^3_2=-x_1 g^3(r)$, $A^3_\mu=0, \mu>2$. It should be
noticed that this example is abelian, since
$F^a_{\mu\nu}$ has only linear terms in $A^a_\mu$. A similar result can be
found in \cite{Capri:2012ev}.

It is important to show solutions of \equ{sol1dd} which are finite for all $r$.
If we choose $\tau(r)=r^q/(r^p+\alpha)$ where $\alpha,p,q$ are positive real
numbers, we have
\begin{eqnarray}\label{g32D}
g^3(r)&=&\frac{\alpha^2 (q-1) (1-q-D)}{r^2
\left(\alpha+r^p\right)^2}+\nonumber\\
&&+\frac{\alpha \left(p^2-D
(2q-2-p)-2 p (q-1)-2 (q-1)^2\right) r^{p}}{r^2
\left(\alpha+r^p\right)^2}+\nonumber\\
&&-\frac{(q-1-p) (q-1-p+D) r^{2 p}}{r^2
\left(\alpha+r^p\right)^2}.
\end{eqnarray}
The function $g^3(r)$ is non-singular at $r=0$ if $p\ge2$ and $q=1$. At larger
values of $r$, we get asymptotically that $g^3(r)\sim r^{-2}$ if $ p\ne D$. This
is not a good behavior for this function as it leads to $A^a_\mu \sim r^{-1}$ as
$r\rightarrow\infty$, with an infinite Hilbert norm  $||A||^2=
\int d^Dx A^a_\mu A^a_\mu $, which makes $\int d^Dx F^a_{\mu\nu}
F^a_{\mu\nu}$ also infinite. The allowed alternative is to take $p=D$ together
with $q=1$ in \equ{g32D}. These choices make the first and the last terms in
\equ{g32D} vanish, and the asymptotic behavior of $g^3(r)$ becomes $g^3(r)\sim
r^{-2-D}$. Consequently,  $A^a_\mu \sim
r^{-1-D}$ as $r\rightarrow\infty$. Then, for $p=D$ the Hilbert norm  $||A||^2=
\int d^Dx A^a_\mu A^a_\mu $ is finite and consequently $\int d^Dx F^a_{\mu\nu}
F^a_{\mu\nu}$ is also finite.

 In order to obtain  solutions for $D>2$, where we have a nonvanishing nonlinear term of $F^a_{\mu\nu}$,   we propose to rewrite eq.
\equ{es1} in a matrix form
\be{mtes1}
\hat{S}_{1D}B_i=R_{ij}C_j,
\ee
where we ordered the $\tau^a_i$ and $M^a_{\mu\nu}$ as $\tau^1_1=B_1,
\tau^1_2=B_2\cdots,\tau^{(N^2-1)}_D=B_{(N^2-1)D}$ and
$M^1_{12}=C_1,M^1_{13}=C_2,\cdots, M^{N^2-1}_{(D-1)D}=C_{(N^2-1)D(D-1)/2}$. In a
general form
 \begin{eqnarray}\label{ijform}
  \tau^a_\mu&\rightarrow& B_{i(a,\mu)},\quad M^a_{\mu\nu}\rightarrow
sign(\mu,\nu) C_{j(a,\mu,\nu)},\nonumber\\
  i(a,\mu)&=&D(a-1)+\mu,\nonumber\\
  j(a,\mu,\nu)&=&(a-1)D(D-1)/2+\nu-\mu+\nonumber\\
  &+&(\mu-1)(2D-\mu)/2,~\mu<\nu,\\
 j(a,\mu,\nu)&=&(a-1)D(D-1)/2+\mu-\nu\nonumber\\
 &+&(\nu-1)(2D-\nu)/2,~\mu>\nu\nonumber,
 \end{eqnarray}
where $sign(\mu,\nu)=1,~\mu<\nu$ and $sign(\mu,\nu)=-1,~\mu>\nu$. Then, after
these redefinitions, we can establish the expression
$\Sigma_{i(a,\mu)}$ which is just $f^{abc}M^b_{\mu\nu}(r)\tau^c_\nu(r)$
rewritten in terms of the $C_{j(a,\mu,\nu)}$ in \equ{ijform}. Finally, the
$R_{ij}$ matrix can be formally defined as
\be{Rij}
R_{ij}= \frac{\partial\Sigma_i}{\partial C_j} .
\ee

In $D=3$, when we have a system of nine equations with nine unknowns, the
determinant of matrix $R$ in the system given by \equ{mtes1} is
\be{det-su23}
\det R=-2 (\tau^1_3 (\tau^2_2 \tau^3_1-\tau^2_1 \tau^3_2)+\tau^1_2 (-\tau^2_3
\tau^3_1+\tau^2_1 \tau^3_3)+\tau^1_1 (\tau^2_3
\tau^3_2-\tau^2_2 \tau^3_3))^3.
\ee
 The transcription that we have done with eq. \equ{mtes1} now allow us to
understand that we have unique solutions for $M^a_{\mu\nu}$ only if $\det
R\ne0$. We can show an explicitly  solution of \equ{es1} by choosing, as an
example,  $\tau^1_2$, $\tau^2_3$
and $\tau^3_1$ non null and making the remaining vanish. In this case we obtain

\begin{eqnarray}\label{asu2d3}
&& A^1_1=\left(-\frac{\tau^1_2 \hat{S}_{13}\tau^1_2}{2 \tau^2_3
\tau^3_1}+\frac{\hat{S}_{13}\tau^2_3}{2
\tau^3_1}+\frac{\hat{S}_{13}\tau^3_1}{2\tau^2_3}\right) x_3,\nonumber\\
&&A^1_2=0,\nonumber\\
&&A^1_3=\left(\frac{\tau^1_2 \hat{S}_{13}\tau^1_2}{2 \tau^2_3
\tau^3_1}-\frac{\hat{S}_{13}\tau^2_3}{2 \tau^3_1}-\frac{\hat{S}_{13}\tau^3_1}{2
\tau^2_3}\right) x_1,\nonumber\\
&&A^2_1=\left(-\frac{\hat{S}_{13}\tau^1_2}{2 \tau^3_1}+\frac{\tau^2_3
\hat{S}_{13}\tau^2_3}{2 \tau^1_2
\tau^3_1}-\frac{\hat{S}_{13}\tau^3_1}{2 \tau^1_2}\right) x_2,\nonumber\\
&&A^2_2=\left(\frac{\hat{S}_{13}\tau^1_2}{2 \tau^3_1}-\frac{\tau^2_3
\hat{S}_{13}\tau^2_3}{2 \tau^1_2 \tau^3_1}+\frac{\hat{S}_{13}\tau^3_1}{2
\tau^1_2}\right) x_1,\\
&&A^2_3=0,\nonumber\\
&&A^3_1=0,\nonumber\\
&&A^3_2=\left(-\frac{\hat{S}_{13}\tau^1_2}{2
\tau^2_3}-\frac{\hat{S}_{13}\tau^2_3}{2 \tau^1_2}+\frac{\tau^3_1
\hat{S}_{13}\tau^3_1}{2
\tau^1_2 \tau^2_3}\right) x_3,\nonumber\\
&&A^3_3=\left(\frac{\hat{S}_{13}\tau^1_2}{2
\tau^2_3}+\frac{\hat{S}_{13}\tau^2_3}{2 \tau^1_2}-\frac{\tau^3_1
\hat{S}_{13}\tau^3_1}{2 \tau^1_2
\tau^2_3}\right) x_2.\nonumber
\end{eqnarray}
 Finally, a particular non-abelian solution for $D=4$ (in fact, for dimensions
$D>3$), can then be easily obtained by replacing the operator $\hat{S}_{13}$ by
$\hat{S}_{14}$ (or $\hat{S}_{1D}$ in general) in
eqs. \equ{asu2d3} and taking $M^a_{\mu\nu}=0$ for $\mu,\nu>3$, which implies
that $A^a_\mu=0$ for $\mu>3$.
Also, following the same line of reasoning of the abelian solution \equ{g32D},
if we choose $\tau^1_2(r)=r/(r^D+\alpha)$, $\tau^2_3(r)=r/(r^D+\beta)$
and $\tau^3_1(r)=r/(r^D+\gamma)$,  we have a finite Hilbert norm
$||A||^2$ and a finite Yang-Mills action. With the configurations given by \equ{asu2d3} we
have nonvanishing nonlinear term in $F^a_{\mu\nu}$.

\section{$SU(3)$ group}
The eight generators of $SU(3)$  satisfy the Lie algebra
$[\lambda^a,\lambda^b]=2if^{abc}\lambda^c$, where $\lambda^a$  are the Gell-Man
matrices with $a,b,c$ running from 1 to 8. The nonvanishing structure constants
are
shown in the following table
\begin{center}
\begin{tabular}[c]{|l|l|}
\hline
$abc$ & $2f^{abc}$\\
\hline
123 & ~2\\
147 & ~1\\
156 & -1\\
246 & ~1\\
257 & ~1\\
345 & ~1\\
367 & -1\\
458 & $\sqrt{3}$\\
678 & $\sqrt{3}$\\
\hline
\end{tabular}
\end{center}
The Lie-algebra valued gauge field is $A_\mu=A^a_\mu\lambda^a/2$. Let us
consider first the case $D=2$. We have a system of 16 equations and 8 unknowns.
It is a too large system to be explicitly shown in this text. Anyway, it follows
the general structure displayed in eq. \equ{mtes1}. We can construct a solution
by choosing $B_i=0,~i\ne 8,9$ and $B_8=B_9=B(r)$, were $B(r)$ is a regular
function finite for all $r$. In this case we have $C_i=0,~i\ne 3,8$ and
\be{su3sol0}
C_3+\sqrt{3}C_8=2\hat{S}_{12}B(r)/B(r).
\ee
The solution of \equ{su3sol0} is obtained by introducing another regular
function $\rho(r)$ with $C_8=\rho(r)$ and
$C_3=2\hat{S}_{12}B(r)/B(r)-\sqrt{3}\rho(r)$. Using \equ{ijform}, \equ{omega}
and \equ{amu} we arrive at

\begin{eqnarray}\label{su3sol1}
&& A^a_{\mu}=0, ~a\ne3,8,\nonumber\\
&& A^3_1=(2\hat{S}_{12}B(r)/B(r)-\sqrt{3}\rho(r))x_2,\nonumber\\
&& A^3_2=-(2\hat{S}_{12}B(r)/B(r)-\sqrt{3}\rho(r))x_1,\nonumber\\
&& A^8_1=\rho(r)x_2,\\
&& A^8_2=-\rho(r)x_1,\nonumber\\
&& \omega^a=0,~a\ne 4,5,\nonumber\\
&& \omega^4=B(r)x_2/r,\nonumber\\
&&\omega^5=B(r)x_1/r\nonumber.
\end{eqnarray}
 As in the last section, we can develop solutions for $D$ dimensions by
replacing the operator $\hat{S}_{12}$ by $\hat{S}_{1D}$ in \equ{su3sol1} and
restricting $A^a_{\mu}=0$ for $\mu>2$. Interestingly, this type of solution
gives an abelian field strength since $f^{a38}=0$, as it already happened in the
$SU(2)$ for the solutions coming from the 2 dimensional case.

 In order to obtain a field strength with nonlinear term, we need to consider $D=3$ at
least. In this case we have a matrix system of 24 by 24. For obvious reasons, we
do not write it explicitly, but we can follow the general structure given in
\equ{mtes1} as in the previous cases. If we consider $B_3=B_8=f(r)$ and
$B_i=0,~i\ne 3,8$ we obtain

 \begin{eqnarray}\label{nonsu3}
 && C_6=-\hat{S}_{13}f(r)/f(r)\nonumber\\
  && C_1=C_8=f_1(r)\nonumber\\
 && C_{10}=-C_{17}=f_2(r)\nonumber\\
 && C_{11}=C_{16}=f_3(r)\\
 && C_{13}=-C_{20}=f_4(r)\nonumber\\
 && C_{14}=C_{19}=f_5(r)\nonumber\\
 && C_i=0 , \mbox{for other values}.\nonumber
 \end{eqnarray}
where $f_i(r),~i=1,\cdots,5$ are arbitrary functions. Consequently we have

 \begin{eqnarray}\label{gaugesu3}
  && A^1_1=f_1(r)x_2,~A^1_2=-f_1(r)x_1,~A^1_3=0,\nonumber\\
  &&
A^2_1=0,~A^2_2=-\frac{\hat{S}_{13}f(r)x_3}{f(r)},~A^2_3=\frac{\hat{S}_{13}
f(r)x_2}{f(r)},\nonumber\\
  && A^3_1=f_1(r)x_3,~A^3_2=0,~A^3_3=-f_1(r)x_1,\nonumber\\
  && A^4_1=f_3(r)x_2+f_4(r)x_3,~A^4_2=-f_3(r)x_1,~A^4_3=-f_4(r)x_1,\nonumber\\
  && A^5_1=f_2(r)x_2+f_5(r)x_3,~A^5_2=-f_2(r)x_1,~A^5_3=-f_5(r)x_1,\\
  && A^6_1=f_4(r)x_2-f_3(r)x_3,~A^6_2=-f_4(r)x_1,~A^6_3=f_3(r)x_1,\nonumber\\
  && A^7_1=f_5(r)x_2-f_2(r)x_3,~A^7_2=-f_5(r)x_1,~A^7_3=f_2(r)x_1,\nonumber\\
  && A^8_\mu=0.\nonumber
 \end{eqnarray}
 \begin{eqnarray}\label{omegasu3}
  \omega^1=f(r)\frac{x_3}{r},~\omega^3=f(r)\frac{x_2}{r},~ \omega^a=0, ~a\ne
1,3.
 \end{eqnarray}
 As in the last section, we can choose $f(r)=r/(r^3+\alpha)$ and suitable
functions $f_i(r)$ to make the Hilbert norm $||A||^2$ finite. By using the fact that $\int dx^D F(r) x_\nu x_\nu=0$ for an arbitrary radial function $F(r)$ and $\mu\ne\nu$, we can make the choices: $f_i(r)=1/(r^{m_i}+\beta_i)$ with $m_i>5/2$ and $\beta_i>0$. Note that here we have more freedom to obtain $||A||^2$ finite than in the $SU(2)$ case.

 The solutions for  $D=4$ can be obtained explicitly  by replacing the operator $\hat{S}_{13}$ by
$\hat{S}_{14}$ in \equ{gaugesu3} and taking all $A^a_4=0$. In this case,  $f_i(r)=1/(r^{m_i}+\beta_i)$ with $m_i>3$, $\beta_i>0$ and   $f(r)=r/(r^4+\alpha)$.
After making these substitutions and performing some calculations we get
\begin{eqnarray}\label{gauged4su3}
  && A^1_1=\frac{x_2}{r^{m_1}+\beta_1},~A^1_2=-\frac{x_1}{r^{m_1}+\beta_1},~A^1_3=0,\nonumber\\
  && A^2_1=0,~A^2_2=\frac{32\alpha r^2 x_3}{(r^4+\alpha)^2},~A^2_3=-\frac{32\alpha r^2 x_2}{(r^4+\alpha)^2},\nonumber\\
  && A^3_1=\frac{x_3}{r^{m_1}+\beta_1},~A^3_2=0,~A^3_3=-\frac{x_1}{r^{m_1}+\beta_1},\nonumber\\
  && A^4_1=\frac{x_2}{r^{m_3}+\beta_3}+\frac{x_3}{r^{m_4}+\beta_4},~A^4_2=-\frac{x_1}{r^{m_3}+\beta_3},~A^4_3=-\frac{x_1}{r^{m_4}+\beta_4},\nonumber\\
  && A^5_1=\frac{x_2}{r^{m_2}+\beta_2}+\frac{x_3}{r^{m_5}+\beta_5},~A^5_2=-\frac{x_1}{r^{m_2}+\beta_2},~A^5_3=-\frac{x_1}{r^{m_5}+\beta_5},\\
  && A^6_1=\frac{x_2}{r^{m_4}+\beta_4}-\frac{x_3}{r^{m_3}+\beta_3},~A^6_2=-\frac{x_1}{r^{m_4}+\beta_4},~A^6_3=\frac{x_1}{r^{m_3}+\beta_3},\nonumber\\
  && A^7_1=\frac{x_2}{r^{m_5}+\beta_5}-\frac{x_3}{r^{m_2}+\beta_2},~A^7_2=-\frac{x_1}{r^{m_5}+\beta_5},~A^7_3=\frac{x_1}{r^{m_2}+\beta_2},\nonumber\\
  && A^8_\mu=0,\nonumber
 \end{eqnarray}
 \begin{eqnarray}\label{omegad4su3}
  \omega^1=\frac{x_3}{r^4+\alpha},~\omega^3=\frac{x_2}{r^4+\alpha},~ \omega^a=0, ~a\ne
1,3.
 \end{eqnarray}
The configurations given by \equ{gauged4su3} provide a nonvanishing nonlinear term in $F^a_{\mu\nu}$ and a finite Hilbert norm $||A||^2$.
 \section{Conclusion}
The final conclusion is that there are in fact non-abelian solutions of the
Gribov problem for $SU(3)$ in $D=4$. They can be easily constructed from the 3
dimensional solution given by \equ{gaugesu3} and \equ{omegasu3} following the
same procedure done in $SU(2)$ case by replacing the operator $\hat{S}_{13}$ by
$\hat{S}_{14}$ in \equ{gaugesu3} and taking all $A^a_\mu=0$ for $\mu=4$. This
means that the Landau gauge does not really eliminate all gauge copies from the
field space of a $SU(3)$ theory in $D=4$.
\section*{Acknowledgements}
The Conselho Nacional de Desenvolvimento Cient\'{\i}ico e Tecnol\'{o}gico CNPq-
Brazil, Funda\c{c}\~{a}o de Amparo a Pesquisa do Estado do Rio de Janeiro
(Faperj) and the SR2-UERJ are acknowledged for the financial support. We wish to thank Dr. S. P. Sorella for
helpful discussions and reading of this manuscript.

R. R. Landim dedicates this paper to the memory of his wife Isabel Mara.
 \appendix
\section{A few remarks on the  Laplace equation in the
Euclidean space}
Let us consider the  angular operator in the Euclidean space
\be{ang-A}
\hat{L}_{\mu\nu}=x_\mu\pa_\nu-x_\nu\pa_\mu.
\ee
From this we obtain
$$\hat{L}^2=\hat{L}_{\mu\nu}\hat{L}_{\mu\nu}
/2=(x_\mu\pa_\nu-x_\nu\pa_\mu)(x_\mu\pa_\nu-x_\nu\pa_\mu)/2=(r^2\pa^2+(1-D))
\frac{\pa}{\pa r}-x_\mu x_\nu\pa_\mu\pa_\nu),
$$
where we use the notation $\pa_\mu\pa_\mu=\pa^2$. By using the fact that
$x_\mu\pa_\mu=r\frac{\pa}{\pa r}$ we also have
\be{lap-A}
\pa^2=\frac{1}{r^{D-1}}\frac{\pa}{\pa r}(r^{D-1}\frac{\pa}{\pa
r})+\frac{1}{r^2}\hat{L}^2.
\ee
The solution of the Laplace equation, that is radial dependent only, satisfies
the radial equation
\be{radial-A}
\frac{1}{r^{D-1}}\frac{d}{dr}(r^{D-1}\frac{d\Phi(r)}{dr})=0,
\ee
and is given by
\begin{eqnarray}\label{sol1-A}
\Phi(r)=\frac{a_0}{r^{D-2}}+b_0,\quad D\ne2, r\ne0.\\
\Phi(r)=a_0\ln r+b_0, \quad D=2, r\ne0.
\end{eqnarray}
That is  for $D>2$,
$$
\pa^2\frac{1}{r^{D-2}}=0, \quad r\ne0.
$$
Since  $\pa^2$ commutes with$\pa_\mu$ we have
\be{d2}
\pa^2\pa_{\mu_1}\pa_{\mu_2}\cdots\pa_{\mu_l} \frac{1}{r^{D-2}}=0, \quad r\ne0.
\ee
Then
$$
\pa_{\mu_1}\pa_{\mu_2}\cdots\pa_{\mu_l} \frac{1}{r^{D-2}},
$$
is a solution of the  Laplace equation.
We now show that
$$
r^{l+D-2}\pa_{\mu_1}\pa_{\mu_2}\cdots\pa_{\mu_l} \frac{1}{r^{D-2}},
$$
is radial independent. Let us introduce a scaling  $r=ar'$ and $x_\mu=ax'_\mu$.
Since $\pa_\mu=a^{-1}\pa'_\mu$, we obtain
$$
r^{l+D-2}\pa_{\mu_1}\pa_{\mu_2}\cdots\pa_{\mu_l}
\frac{1}{r^{D-2}}=r'^{l+D-2}\pa'_{\mu_1}\pa'_{\mu_2}\cdots\pa'_{\mu_l}
\frac{1}{r'^{D-2}}.
$$
This means that $r^{l+D-2}\pa_{\mu_1}\pa_{\mu_2}\cdots\pa_{\mu_l}
\frac{1}{r^{D-2}}=F(x_1,x_2,\cdots x_D)$ is a function such that
$$
F(x_1,x_2,\cdots x_D)=F(ax_1,ax_2\cdots ax_D).
$$
Making the derivative with respect to
$a$ and taking  $a=1$ we establish that
$$
x_\mu\pa_\mu F=r\frac{\pa F}{\pa r}=0.
$$
This implies that $\frac{\pa F}{\pa r}=0$\footnote{For $D=2,  \ln(ar')=\ln
a+\ln r'$ and the function which is radial free will be
$r^l\pa_{\mu_1}\cdots\pa_{\mu_l}\ln r$.}.
We define the tensors $Q_{\mu_1\mu_2\cdots \mu_l}(\Omega)$ as
\begin{eqnarray}\label{tensores-A}
Q_{\mu_1\mu_2\cdots
\mu_l}(\Omega)=(-1)^l\frac{r^{l+D-2}}{l!(D-2)}\pa_{\mu_1}\pa_{\mu_2}\cdots\pa_{
\mu_l} \frac{1}{r^{D-2}}, \quad D>2,\\
Q_{\mu_1\mu_2\cdots
\mu_l}=(-1)^{(l+1)}\frac{r^l}{l!}\pa_{\mu_1}\pa_{\mu_2}\cdots\pa_{\mu_l}\ln r,
\quad D=2.
\end{eqnarray}
From \equ{d2}, we have
$$
\frac{1}{r^{D-1}}\frac{\pa}{\pa r}(r^{D-1}\frac{\pa}{\pa r}
\frac{1}{r^{l+D-2}})Q_{\mu_1\mu_2\cdots
\mu_l}(\Omega)+\frac{1}{r^{l+d}}\hat{L}^2 Q_{\mu_1\mu_2\cdots \mu_l}(\Omega)=0,
\quad r\ne0.
$$
Then, these tensors satisfy the eigenvalue equation

\be{auto-A}
\hat{L}^2 Q_{\mu_1\mu_2\cdots \mu_l}(\Omega)=-l(l+D-2)Q_{\mu_1\mu_2\cdots
\mu_l}(\Omega),
\ee
and using \equ{d2} once more, we see that
$$
 Q_{\nu\nu\mu_3\cdots \mu_l}(\Omega)=r^{l+D-2}\pa^2\pa_{\mu_3}\cdots\pa_{\mu_l}
\frac{1}{r^{D-2}}=0,
 $$
i.e., the $Q_{\mu_1\mu_2\cdots \mu_l}(\Omega)$ are traceless.
In this way, the number of independent components of these tensors is
\be{comp-A}
N_c=\left(\begin{array}{c}
        D+l-1\\
        l
       \end{array}
       \right)-\left(\begin{array}{c}
        D+l-3\\
        l-2
       \end{array}
       \right)=\frac{(D+l-3)!(2l+D-2)}{l!(D-2)!}
\ee
For example, in $D=2$ we have $N_c=2$, to $D=3$ , $N_c=2l+1$, $D=4$
 $N_c=(l+1)^2$, etc.

We show bellow some of these tensors in cartesian coordinates
\begin{eqnarray}
&&Q_\mu=x_\mu/r,\\
&&Q_{\mu\nu}=\frac{1}{2}\frac{(Dx_\mu x_\nu-r^2\delta_{\mu\nu})}{r^2}.
\end{eqnarray}
The action of the operator $\hat{O}^a=M^a_{\mu\nu}(r)x_\nu\pa_\mu$ in the
tensors
$Q$ follows from the algebraic identity
\be{al-id}
M^a_{\mu\nu}(r)x_\nu\pa_\mu\frac{1}{r}=0.
\ee
Deriving \equ{al-id} $l$ times with respect of $x_{\mu_i}$ we obtain
\be{OQ-A}
\hat{O}^aQ_{\mu_1\cdots \mu_l}(\Omega)=M^a_{\mu_1\tau}(r)Q_{\tau\cdots
\mu_l}(\Omega)+\cdots+ M^a_{\mu_l\tau}(r)Q_{\mu_1\cdots \mu_l}(\Omega).
\ee


\begin{thebibliography}{99}

\bibitem{Henyey:1978qd}
  F.~S.~Henyey,
  Phys.\ Rev.\  {\bf D20}, 1460 (1979).

\bibitem{Gribov:1977wm}
  V.~N.~Gribov,
  Nucl.\ Phys.\  B {\bf 139}, 1 (1978).

\bibitem{Singer:1978dk}
  I.~M.~Singer,
  Commun.\ Math.\ Phys.\  {\bf 60}, 7 (1978).

\bibitem{Zwanziger:1988jt}
 D.~Zwanziger,
Nucl.\ Phys.\  B {\bf 321}, 591 (1989).


\bibitem{Zwanziger:1989mf}
  D.~Zwanziger,
  Nucl.\ Phys.\  B {\bf 323}, 513 (1989).






%
%


%



%



%
%
%




\bibitem{Capri:2010hb}
M.~Capri, A.~Gomez, M.~Guimaraes, V.~Lemes, S.~Sorella, {\em et al.},
  {{\em Phys.Rev.} {\bf D82}
  (2010)  105019}, {\tt arXiv:1009.4135
  [hep-th]}.

\bibitem{Gomez:2009tj}
A.~J. Gomez, M.~S. Guimaraes, R.~F. Sobreiro, and S.~P. Sorella, {{\em Phys.
  Lett.} {\bf B683} (2010)  217--221},
{{\tt arXiv:0910.3596 [hep-th]}}.

\bibitem{Dudal:2010hj}
D.~Dudal and N.~Vandersickel, {{\tt arXiv:1010.3927 [hep-th]}}.

\bibitem{Sorella:2009vt}
S.~P. Sorella, {{\em Phys. Rev.} {\bf
  D80} (2009)  025013}, {{\tt arXiv:0905.1010 [hep-th]}}.
\bibitem{Dudal:2008sp}
  D.~Dudal, J.~A.~Gracey, S.~P.~Sorella, N.~Vandersickel, H.~Verschelde,
  Phys.\ Rev.\  {\bf D78}, 065047 (2008),
 {\tt arXiv:0806.4348 [hep-th]}.

\bibitem{Dudal:2010cd}
  D.~Dudal, M.~S.~Guimaraes and S.~P.~Sorella,
  Phys.\ Rev.\ Lett.\  {\bf 106}, 062003 (2011),
  {\tt arXiv:1010.3638 [hep-th]}.

\bibitem{Vilar:2011sp}
  L.~C.~Q.~Vilar, O.~S.~Ventura and V.~E.~R.~Lemes,
  Phys.\ Rev.\ D {\bf 84}, 105026 (2011),
  {\tt arXiv:1108.3305 [hep-th]}.
\bibitem{Cucchieri:2011ig}
  A.~Cucchieri, D.~Dudal, T.~Mendes and N.~Vandersickel,
  Phys.\ Rev.\ D {\bf 85}, 094513 (2012),
  {\tt arXiv:1111.2327 [hep-lat]}.








\bibitem{Guimaraes:2011sf}
  M.~S.~Guimaraes and S.~P.~Sorella,
  J.\ Math.\ Phys.\ {\bf 52}, 092302 (2011),
  {\tt arXiv:1106.3944 [hep-th]}.
\bibitem{Capri:2012ev}
  M.~A.~L.~Capri, M.~S.~Guimaraes, S.~P.~Sorella and D.~G.~Tedesco,
  Eur.\ Phys.\ J.\ C {\bf 72}, 1939 (2012),
  {\tt arXiv:1201.2445 [hep-th]}.


%
%
%
%





%
%
%
%
%







\end{thebibliography}
\end{document}